\documentclass[showpacs,twocolumn,aps,prd]{revtex4-1}
\usepackage{graphicx}
\usepackage{dcolumn}
\usepackage{bm}
\usepackage{amsmath}
\usepackage{amssymb}
\usepackage[usenames,dvipsnames]{xcolor}
\usepackage{color}
\usepackage[colorlinks,plainpages=false]{hyperref}

\begin{document}
  \title{Properties of strange quark matter and evolution of strange quark stars}
\author{Huai-Min~Chen,$^1$}
\email{chenhuaimin@wuyiu.edu.cn}
\author{Cheng-Jun~Xia$^{2}$}
\author{Ren-Xin~Xu$^{3,4}$}
\author{Guang-Xiong~Peng$^{5,6,7}$}

\affiliation{%
    $^1$\mbox{School of Mechanical and Electrical Engineering, Wuyi University, Wuyishan 354300, China}\\
	$^2$\mbox{Center for Gravitation and Cosmology, College of Physical Science and Technology,} \\
    {Yangzhou University, Yangzhou 225009, China}\\
    $^3$\mbox{School of Physics, Peking University, Beijing 100871, China}\\
    $^4$\mbox{Kavli Institute for Astronomy and Astrophysics, Peking University, Beijing 100871, China}\\
    $^5$\mbox{School of Nuclear Science and Technology, University of Chinese Academy of Sciences,} \\
    {Beijing 100049, China}\\
	$^6$\mbox{Theoretical Physics Center for Science Facilities, Institute of High Energy Physics, P.O. Box 918,}  \\
    {Beijing 100049, China}\\
	$^7$\mbox{Synergetic Innovation Center for Quantum Effects and Application, Hunan Normal University,}  \\
    {Changsha 410081, China}  }

\begin{abstract}
In this work, we study the properties of strange quark matter and reveal the evolution process of strange quark stars employing a self-consistent thermodynamic treatment. A comprehensive and reliable thermodynamic basis for the study of the dynamic evolution from proto-strange quark stars to stable strange quark stars at zero temperature is provided. The relative abundance of particles, equation of state, temperature, and mass-radius relationship at each stage of stellar evolution are discussed. The cold strange quark star is consistent with the observational masses and radii of HESS J1731-347, PSR J1231-1411, PSR J0030+0451, PSR J0348+0432, and PSR J0740+6620, which are difficult to explain with the standard neutron star model. A schematic diagram is also provided, illustrating the state of different stages along the evolution of stars at fixed baryon mass.

\end{abstract}

\maketitle
\section{Introduction}
\label{intro}
Strange quark matter (SQM) is a new form of matter predicted by quantum chromodynamics (QCD), composed of approximately equal numbers of $u$, $d$, and $s$ quarks and a small number of electrons. In the 1970s and 1980s, Witten and Bodmer hypothesized that strange quark matter might be the true ground state of strongly interacting matter~\cite{Bodmer1971_PRD4-1601,Witten1984_PRD30-272}. If this is the case, then there must exist objects composed of SQM, such as strange quark stars (SQS)~\cite{Xia2014_PRD89-105027,Sedaghat2025_EPJC85-283}, strangelets~\cite{Chen2024_PRD109-054013,Lugones2023_PRD107-043025}, meteorlike compact ultradense objects~\cite{Rafelski2012_APPB43-2251,Rafelski2013_PRL110-111102}, and nuclearites~\cite{Piotrowski2020_PRL125-091101,ANTARES2023_JCAP01-012}, etc., which contain baryon numbers ranging from a few to approximately $10^{57}$. For small lumps of SQM with fewer baryons ($A\leq10^7$), they are generally referred to as strangelets~\cite{Chen2024_PRD109-054031,You2024_PRD109-034003}. When the size of dense objects composed of SQM reaches a macroscopic scale and becomes a compact star observable by astronomy ($A\approx10^{57}$), such compact stars are called strange quark stars and represent one of the candidates for pulsars~\cite{Lugones2024_Universe10-233}. Most subsequent theoretical studies have shown that SQM can exist stably and is more stable than ordinary nuclear matter~\cite{Farhi1984_PRD30-2379,Chen2022_CPC46-055102}. Furthermore, it was found that SQM is consistent with experiments in predicting the phenomena of some strong interaction processes~\cite{Xu2017_PRD96-063016,Zhang2021_PRD103-103021}. Therefore, strange quark stars and strangelets consisting of SQM may exist widely in the universe and are expected to be observed.

It is generally believed that strange quark stars may be directly generated during type II supernova explosions if strange quark matter is absolutely stable. For example, after a massive star undergoes a supernova explosion, the core of the star is likely to further collapse and form a strange quark star~\cite{Cheng1998_JMPD7-139}. Meanwhile, ordinary neutron stars may also undergo phase transitions and form strange quark stars. For example, after the annihilation of weakly interacting massive particles (WIMPs) accreted by a regular neutron star, strangelets are generated inside the neutron star, which quickly transform the neutron star into a strange quark star or hybrid star~\cite{Perez2010_PRL105-141101}. In fact, as long as there are trace amounts of strangelets in the universe, all neutron stars have the potential to be transformed into strange quark stars~\cite{Alcock1986_AJ310-261,Olinto1987_PLB192-71}. This conversion takes only a few milliseconds to transform a neutron star into a strange quark star~\cite{Herzog2011_PRD84-083002}.

The possibility of a quark matter core inside a neutron star was first explored by Ivanenko and Kurdgelaidze in 1965~\cite{Ivanenko1965_A1-251,Ivanenko1969_LNC2-13}. Later, Itoh further developed the notion that dense stars are actually quark stars composed entirely of strange quark matter~\cite{Itoh1970_PTP44-291}. Subsequently, many works investigated the dense matter and structure of dense stars~\cite{Issifu2024_EPJC84-463,Song2025_PRD111-063018}. Recently, by considering the isentropic stage of strange quark star evolution, we discussed the structure of pure strange quark stars in the baryon density-dependent quark mass model, and identified a massive proto-strange quark star with a maximum mass exceeding twice the solar mass, which provided further insight into the early stages of evolution of a strange quark star~\cite{Chen2022_CPC46-055102}. In a theoretical interpretation of observational data from the neutron star EXO 0748-676, Alford's study showed that the presence of quark matter in EXO 0748-676 could not be ruled out~\cite{Alford2007_Nat445-E7}. Drake and Slane's research suggests that two hypothetical neutron stars, RXJ 1856.5-3754 in Corona Australis and 3C58 in Cassiopeia, are too small and dense to be neutron stars, indicating the possibility of quark stars~\cite{Drake2002_AJ572-996,Slane2002_AJL571-L45,Prakapenia2026_PRD114-043029}. In Ju's work, they studied the mass-radius relation of HESS J1731-347 based on the MIT bag model, and the results support the hypothesis that HESS J1731-347 is a quark star~\cite{Ju2025_EPJC85-40}.
Based on perturbative QCD corrections and color superconductivity, Zhang and Mann's work indicates that currently observed compact stars (including the recently reported GW190814's secondary component) can be quark stars composed of interacting quark matter~\cite{Zhang2021_PRD103-063018}.
In addition, no neutron star has been found for SN 1987A after core-collapse, and studies have suggested that the dense remnant of SN 1987A may be a strange quark star or strangeon star~\cite{Chan2009_AJ695-732,Yuan2017_RAA17-92}. The brightest supernova in history, SN 2006gy, exhibited exceptional brightness, being several hundred times brighter than a typical Type II supernova. This abnormally bright supernova indicates that there is a large amount of extra energy, leading to the suspicion that some kind of quark star may have been born there~\cite{Ouyed2012_MNRAS423-1652}.

As one of the candidates for pulsars, strange quark stars have been studied extensively. In most cases, the theoretical description of strange quark stars is similar to that of neutron stars, i.e., thermodynamic conditions (e.g., lepton fraction, entropy per baryon, and neutrino chemical potential) are used to study the evolution of strange quark stars~\cite{Janka2012_PTEP2012-01A309,Fischer2012_PAN75-613}. In addition to the baryon density-dependent quark mass model and MIT bag model mentioned above, other phenomenological models have also been applied to the study of strange quark stars, including the Nambu-Jona-Lasinio model~\cite{Chu2024_PRD110-123031,Sedaghat2024_EPJC84-171}, quasiparticle model~\cite{Wen2005_PRC72-015204,Yang2025_EPJC85-426}, strangeon star model~\cite{Xu2003_AJ596-L59}, perturbative model~\cite{Peng2005_EL72-69,Xu2015_PRD92-025025,Xu2017_PRD96-063016}, and so on.

In this paper, we apply the baryon density-dependent quark mass model to study the evolution of a strange quark star from a proto-strange quark star at birth to a stable strange quark star at zero temperature by considering $\beta$-equilibrium matter in the neutrino-trapped and neutrino-transparent regimes during the deleptonization and cooling processes of the strange quark star. The evolution of proto-strange quark stars involves multiple stages, each with significantly different thermodynamic conditions. By modeling the evolution in stages, the evolution of strange quark stars can be described more accurately. It is important to note that in any case the modeling results must satisfy thermodynamic self-consistency, which we have discussed in detail in this paper. By applying a self-consistent thermodynamic treatment and requiring the strange quark matter to satisfy the corresponding local electrical neutrality and chemical equilibrium conditions, the relationships between the various thermodynamic quantities of the strange quark matter are derived in a self-consistent way.

The present paper is organized as follows. In Sec.~\ref{Therm}, we discuss the thermodynamic treatment of SQM with baryon density-dependent quark mass model. In Sec.~\ref{sec:mass}, we give the quark and gluon mass scaling at finite and zero temperatures, where the equivalent mass of quarks takes into account confinement and perturbation interactions. In Sec.~\ref{SQMProp}, we present the numerical results for the properties of SQM. In Sec.~\ref{SQSProp}, the evolution of a strange quark star are discussed. Finally, a short summary is given in Sec.~\ref{Summ}.

\section{Self-consistent thermodynamic treatment}
\label{Therm}
The most complex and important issue in the equivparticle model is thermodynamic self-consistency, for which various thermodynamic treatments have been developed based on different considerations~\cite{Wen2005_PRC72-015204,Benvenuto1995_PRD51-1989,Peng2000_PRC62-025801}. To introduce the interaction between quarks in a free particle system, the constant quark mass in the thermodynamic expression of the free particle system is replaced with an equivalent mass $m_{i}=m_{i}(n_u,n_d,n_s,T)$ that varies with the environment, altering the relationship between various thermodynamic quantities derived from the original free particle system, therefore it is necessary to be treated with caution. In this paper, the free energy density $F$ is taken as the characteristic thermodynamic function, and the expressions of each thermodynamic quantity are obtained by self-consistent thermodynamic treatment. The free energy density adopts the same form as that of the free particle system, but the mass in the expression is replaced by an equivalent mass that depends on density and temperature. Since the objects of interest are bulk strange quark matter and strange quark stars, the independent state parameters are particle number density $n_{i}$ and temperature $T$. The free energy density is expressed as
\begin{eqnarray}\label{1}
F &=& F(T,\{n_{i}\},\{m_{i}\})\nonumber\\
  &=& \Omega_{0}(T,\{\mu_{i}^{*}\},\{m_{i}\})+\sum_{i}\mu_{i}^{*}n_{i}.
\end{eqnarray}
where $\mu_i^*$ is the effective chemical potential of particle type $i$, and $\Omega_{0}$ is the thermodynamic potential density of a system comprised of free particles, which should be considered as an intermediate variable rather than a real one. The corresponding differential equation of Eq.~(\ref{1}) then becomes
\begin{eqnarray}\label{2}
\mathrm{d}F &=& \frac{\partial\Omega_{0}}{\partial T}\mathrm{d}T+\sum_{i}\mu_{i}^{*}\mathrm{d}n_{i}+\sum_{i}\frac{\partial\Omega_{0}}{\partial m_{i}}\mathrm{d}m_{i},
\end{eqnarray}
where the differential of $m_{i}$ is as follows:
\begin{eqnarray}\label{3}
\mathrm{d}m_{i} = \sum_{j}\frac{\partial m_{i}}{\partial n_{j}}\mathrm{d}n_{j}+\frac{\partial m_{i}}{\partial T}\mathrm{d}T.
\end{eqnarray}
Then, we have
\begin{eqnarray}\label{4}
\mathrm{d}F &=& \left[
      \frac{\partial \Omega_0}{\partial T}
      + \sum_i \frac{\partial \Omega_0}{\partial m_i} \frac{\partial m_i}{\partial T}
\right]\mbox{d}T  \nonumber \\
&&
+
\sum_i \left[ \mu_i^* +  \sum_j\frac{\partial \Omega_0}{\partial m_j}\frac{\partial m_j}{\partial n_i}
\right] \mbox{d} n_i.
\end{eqnarray}

The fundamental thermodynamic differential relation of free energy is
\begin{equation}\label{5}
\mathrm{d}\overline{F}=-\overline{S}\mathrm{d}T-P\mathrm{d}V+\sum_{i}\mu_{i}\mathrm{d}{N_{i}},
\end{equation}
where $\overline{F}$, $\overline{S}$, $N_{i}$, $P$, $V$, and $\mu_{i}$ represent the free energy, entropy, particle number, pressure, volume, and chemical potential of the system, respectively. For a uniform system, we can define the free energy density $F =\overline{F}/V $, entropy density $S =\overline{S}/V$, and particle number density $n_{i} =N_{i}/V$. Substituting them into Eq.~(\ref{5}), we have
\begin{eqnarray}\label{6}
\mathrm{d}F
&=&-S\mathrm{d}T+\sum_{i}\mu_{i}\mathrm{d}n_{i} \nonumber\\
& &{}+\left(-P-F+\sum_{i}\mu_{i}n_{i}\right)\frac{\mathrm{d}V}{V}.
\end{eqnarray}

Comparing Eqs.~(\ref{4}) and ~(\ref{6}), we obtain the following thermodynamic relationships,
\begin{eqnarray}
S &=& -\frac{\partial\Omega_{0}}{\partial T}-\sum_{i}\frac{\partial\Omega_{0}}{\partial m_{i}}\frac{\partial m_{i}}{\partial T}, \label{7}
\end{eqnarray}
\begin{eqnarray}
\mu_{i} &=& \mu_i^* +   \sum_j\frac{\partial \Omega_0}{\partial m_j}\frac{\partial m_j}{\partial n_i}, \label{8}
\end{eqnarray}
\begin{eqnarray}
P &=&  -F+\sum_i \mu_{i} n_i. \label{9}
\end{eqnarray}
Based on $F=\Omega_0+\sum {\mu_{i}^{*}n_{i}}=\Omega+\sum {\mu_{i}n_{i}}$, the relationship between intermediate variables $\Omega_{0}$ and true thermodynamic potential $\Omega$ is
\begin{eqnarray}\label{10}
    \Omega=\Omega_0 -n_b\sum_j\frac{\partial \Omega_0}{\partial m_j}\frac{\partial m_j}{\partial n_b}.
\end{eqnarray}
Substituting Eqs.~(\ref{1}) and ~(\ref{10}) into Eq.~(\ref{9}),
\begin{eqnarray}\label{11}
        P =  -\Omega_{0}+n_b\sum_j\frac{\partial \Omega_0}{\partial m_j}\frac{\partial m_j}{\partial n_b}=-\Omega,
\end{eqnarray}
which demonstrates the consistency of the entire process.

The energy density is given by
\begin{eqnarray}\label{12}
        E=F+TS=\Omega_{0}+\sum_{i}{\mu_{i}^{*}n_{i}}+TS.
\end{eqnarray}
Substituting Eq.~(\ref{9}) into Eq.~(\ref{12}), we have
\begin{eqnarray}\label{13}
        E=-P+\sum_{i}{\mu_{i}n_{i}}+TS,
\end{eqnarray}
so the Euler equation for bulk strange quark matter is satisfied.

According to $F(T,\{n_{i}\},\{m_{i}\})=\Omega_0(T,\{\mu_{i}^{*}\},\{m_{i}\})+\sum {\mu_{i}^{*}n_{i}}$, we have $\partial F/\partial \mu_{i}^{*}=\partial \Omega_{0}/\partial \mu_{i}^{*} + n_{i}=0$. The particle number density is
\begin{eqnarray}\label{14}
n_{i} = -\frac{\partial\Omega_{0}}{\partial \mu_{i}^{*}}.
\end{eqnarray}

\section{Density and/or temperature dependent particle masses}
\label{sec:mass}
The main idea of the model is to simulate the interactions between quarks by using baryon density-dependent quark masses. The corresponding quarks behave as free particles with equivalent masses. This idea was initially proposed by Fowler and further developed by Chakrabarty~\cite{Fowler1981_ZPC9-271,Chakrabarty1993_PRD48-1409}, namely
\begin{equation}\label{15}
m_i=m_{i0}+\frac{B}{3n_\mathrm{b}},
\end{equation}
where $m_{i0}$, $B$, and $n_{b}$ are the current mass ($m_{u0}=2.16\ \mathrm{MeV}$, $m_{d0}=4.70\ \mathrm{MeV}$, and $m_{s0}=93.5\ \mathrm{MeV}$)~\cite{PDG2024_PRD110-030001}, bag constant, and baryon number density.

Afterwards, the temperature contribution is taken into account in the mass scale~\cite{Zhang2002_PRC65-035202}, i.e.,
\begin{equation}\label{16}
m_i=m_{i0}+\frac{B}{3n_\mathrm{b}}\left[1-\left(\frac{T}{T_{c}}\right)^{2}\right],
\end{equation}
where $T_c$ is the critical temperature for the transition of hadronic matter into QGP.

In order to better reflect the interactions between quarks, Peng et al. derived a scaling inversely proportional to the cubic root of the baryon number density based on in-medium chiral condensates and linear confinement~\cite{Peng2000_PRC62-025801,Peng1999_PRC61-015201}, i.e.,
\begin{equation}\label{17}
m_{i}=m_{i0}+ \frac{D}{n_\mathrm{b}^{1/3}},
\end{equation}
where $D$ represents the strength of the confinement interaction. Using a similar derivation, the scaling was extended to finite temperature as~\cite{Wen2005_PRC72-015204}
\begin{equation}\label{18}
m_i =m_{i0}+\frac{D}{n_\mathrm{b}^{1/3}}\left[1-\frac{8T}{\lambda T_{c}}\mathrm{exp}\left(-\frac{\lambda T_{c}}{T}\right)\right],
\end{equation}
where the constant $\lambda=1.6$. $T_{c}=175\ \mathrm{MeV}$ is the critical temperature at the deconfinement phase transition.

Furthermore, considering the contribution of one-gluon-exchange and perturbation interactions, the scaling was obtained at zero temperature~\cite{Xia2014_PRD89-105027,Chen2012_CPC36-947}, i.e.,
\begin{equation}\label{19}
m_i =m_{i0}+\frac{D}{n_\mathrm{b}^{1/3}}+Cn_\mathrm{b}^{1/3},
\end{equation}
where $C$ represents the strength of one-gluon-exchange and perturbation interactions.

Similarly, taking into account the temperature, the scaling becomes~\cite{Lu2016_NST27-148}
\begin{eqnarray}\label{20}
m_i &=& m_{i0}+\frac{D}{n_\mathrm{b}^{1/3}}\left(1+\frac{8T}{\Lambda}e^{-{\Lambda}/{T}}\right)^{-1}\nonumber\\
             & & {} +Cn_\mathrm{b}^{1/3}\left(1+\frac{8T}{\Lambda}e^{-{\Lambda}/{T}}\right),
\end{eqnarray}
where $\Lambda=\lambda T_{c}=280\ \mathrm{MeV}$. For the above formula, $m_i\rightarrow m_{i0}$ above the critical temperature $T_{c}$, which is consistent with the characteristics of asymptotic freedom.

At finite temperatures, it is necessary to know the effective mass of gluons because their contribution cannot be ignored. Based on Bors\'{a}nyi $\emph{et al.}$'s lattice simulation, which provided 48 pressure values~\cite{Borsanyi2012-JHEP01-138}, we obtained a fitting result for the effective mass of gluons~\cite{Chen2022_PRD105-014011}. Define the scaling temperature as $x= T/T_{c}$. For $T<T_{c}$, the effective mass of gluons is
\begin{equation}\label{21}
  \frac{m_{g}}{T}=\sum_{i}a_{i}x^{i}=a_{0}+a_{1}x+a_{2}x^{2}+a_{3}x^{3},
\end{equation}
where $a_{0}=67.018$, $a_{1}=-189.089$, $a_{2}=212.666$, $a_{3}=-83.605$.
For $T> T_{c}$, the effective mass of gluons is
\begin{equation}\label{22}
  \frac{m_{g}}{T}=\sum_{i}b_{i}\alpha^{i}=b_{0}+b_{1}\alpha+b_{2}\alpha^{2}+b_{3}\alpha^{3},
\end{equation}
where expansion coefficients $b_{0}=0.218$, $b_{1}=3.734$, $b_{2}=-1.160$, $b_{3}=0.274$. $\alpha$ is the strong coupling constant, which is a running coupling that depends on the solution of the renormalization group equation. Recently, we solved the renormalization group equations for QCD couplings by a new mathematical idea and obtained a fast convergent expression for $\alpha$~\cite{Chen2022_IJMPE31-2250016}, and the leading term is
\begin{eqnarray}\label{23}
\alpha=\frac{\beta_0}{\beta_0^2\ln(u/\Lambda)+\beta_1\ln\ln(u/\Lambda)},
\end{eqnarray}
where $\beta_0={11}/{2}-{N_f}/{3}$, $\beta_1={51}/{4}-{19N_f}/{12}$. The number of flavors $N_{f}=3$. The renormalization scale varies linearly with temperature as ${u}/{\Lambda}=c_0+c_1x$, where $c_{0}=1.054$ and $c_{1}=0.479$.

\section{Properties of strange quark matter}\label{SQMProp}
SQM is assumed to consist of quarks, gluons, and leptons at finite temperature, but only quarks and electrons at zero temperature. Based on the baryon density-dependent quark mass model, the contribution of thermodynamic potential density for free particles is
\begin{equation}\label{24}
\Omega_{0}=\Omega_{0}^{+}+\Omega_{0}^{-}+\Omega_{0}^{g}.
\end{equation}
The contributions of particles (+) and antiparticles ($-$) are
\begin{equation}\label{25}
\Omega_{0}^{\pm}=\sum_{i}-\frac{d_{i}T}{2\pi^{2}}\int_{0}^{\infty}\ln\left[1+e^{-(\sqrt{p^{2}+m_{i}^{2}}\mp\mu_{i}^{*})/T}\right]p^{2}\mathrm{d}p,
\end{equation}
where $d_{i}=3(colors)\times2(spins)=6$ for $i=u,d,s$, $d_ {i}=2$ for $i=e,\mu$, and $d_ {i}=1$ for $i=\nu_e,\nu_{\mu}$.

The contribution of gluons is
\begin{equation}\label{26}
\Omega_{0}^{g}=\frac{d_{g}T}{2\pi^{2}}\int_{0}^{\infty}\ln\left[1-e^{-\sqrt{p^{2}+m_{g}^{2}}/T}\right]p^{2}\mathrm{d}p,
\end{equation}
where $d_{g}=8(colors)\times2(spins)=16$.

The number density of particle and anti-particle are
\begin{equation}\label{27}
n_{i}^{\pm}=-\frac{\partial\Omega_{0}^{\pm}}{\partial \mu_{i}^{*}}=\frac{d_{i}}{2 \pi^{2}}\int_{0}^{\infty}\frac{p^{2}\mathrm{d}p}{e^{(\sqrt{p^{2}+m_{i}^{2}}\mp \mu_{i}^{*})/T}+1}.
\end{equation}

At zero temperature, the contribution of thermodynamic potential density for free particles is
\begin{equation}\label{28}
\Omega_0=-\sum_i\frac{d_i}{24\pi^2}
 \left[ \mu_i^{*}\nu_i(\nu_i^2-\frac{3}{2}m_i^2)
 +\frac{3}{2} m_i^4\ln\frac{\mu_i^{*}+\nu_i}{m_i}
 \right],
\end{equation}
where $\nu_i=\sqrt{{\mu_{i}^{*}}^2-m_{i}^2}$ is Fermi momentum of the particles.

The number density of particle and anti-particle at zero temperature are
\begin{equation}\label{29}
n_i=-\frac{\partial\Omega_{0}}{\partial \mu_{i}^{*}}=\frac {d_i\nu_i{}^3}{6\pi^{2}}.
\end{equation}

All other thermodynamic quantities can be obtained by substituting the thermodynamic potential density into Eqs.~(\ref{7})-(\ref{12}).
 \begin{figure}
\centering
\includegraphics[width=8cm]{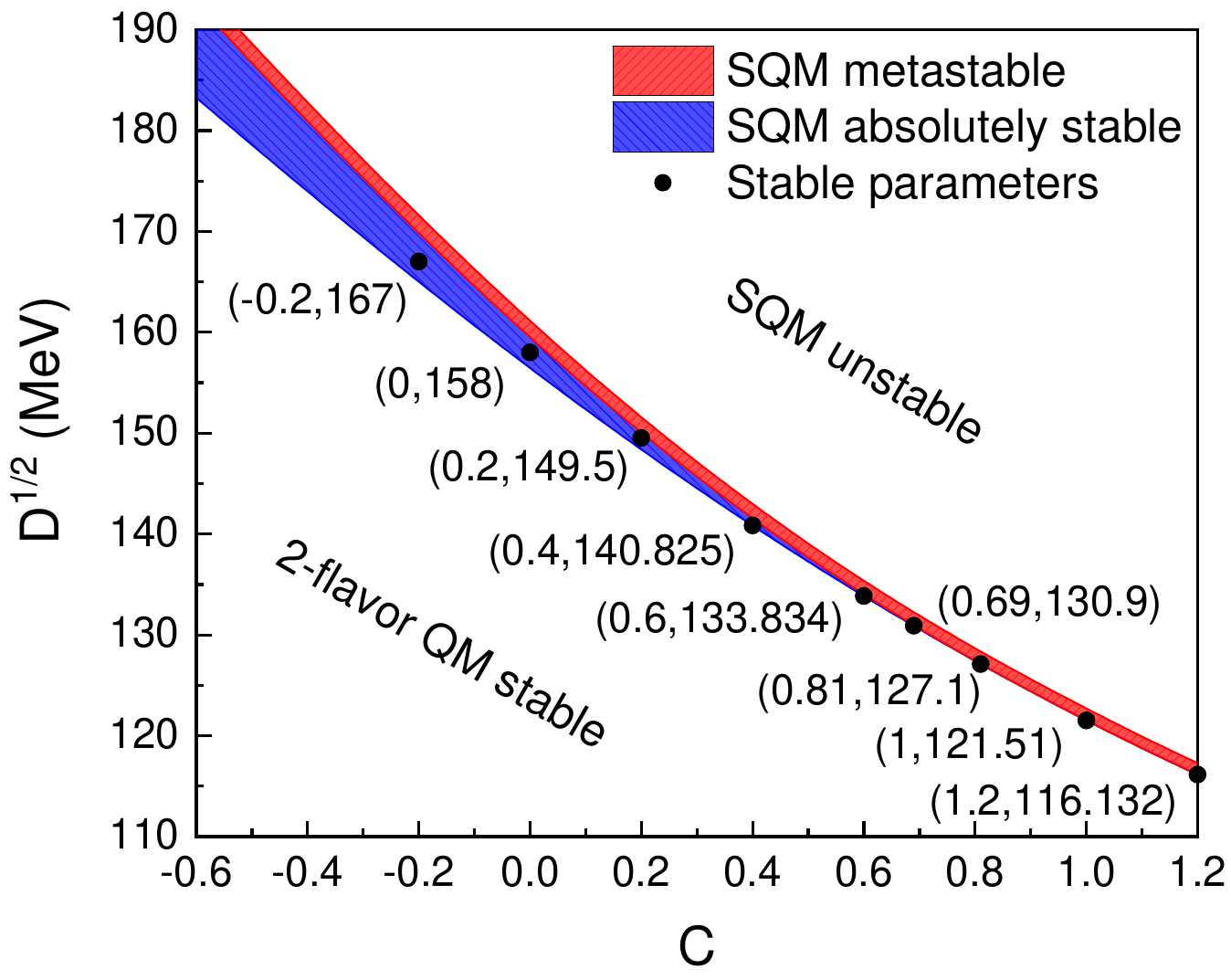}
\caption{The stability window for the baryon density-dependent quark mass model. }
\label{fig1}
\end{figure}

In order to be consistent with conventional nuclear physics and to ensure the stable existence of strange quark stars in the Universe, the confinement parameter $D$ and perturbation parameter $C$ should satisfy the condition that the energy per baryon of strange quark matter is less than $930\ \mathrm{MeV}$ at zero temperature and pressure, while that of $ud$ quark matter is larger than $930\ \mathrm{MeV}$. Furthermore, strange quark matter with a minimum energy per baryon larger than 930 MeV and less than 939 MeV is metastable. Based on the above conditions, we obtain a stability window consisting of the confinement parameter $D^{1/2}$ and perturbation parameter $C$, as shown in Fig.~\ref{fig1}, where the blue and red regions are absolutely stable and metastable SQM, respectively. In this work, we adopt the parameter set ($C$, $\sqrt D$ in MeV) as (0.69, 130.9).

To test the self-consistency of the thermodynamic treatment more intuitively, the thermodynamic benchmark equation is generally defined as~\cite{Peng2000_PRC62-025801}
\begin{equation}\label{30}
P-n^{2}\frac{\mathrm{d}}{\mathrm{d}n}\left(\frac{F}{n}\right)=0.
\end{equation}

\begin{figure}
\centering
\includegraphics[width=8cm]{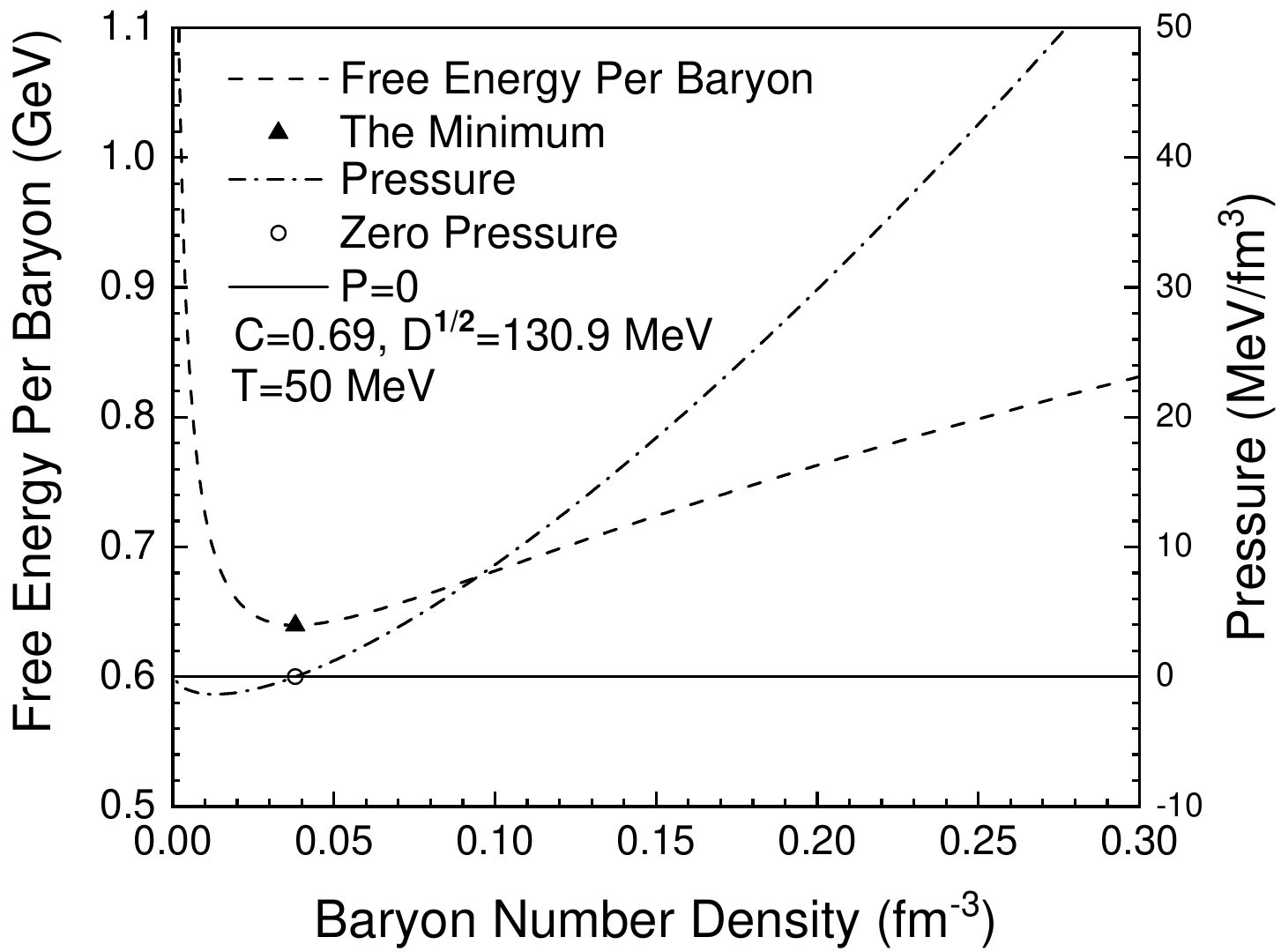}
\caption{The relationship between the free energy per baryon, pressure and density at $T=50\ \mathrm{MeV}$, where $D^{1/ 2}=130.9\ \mathrm{MeV}$ and $C=0.69$. }
\label{fig2}
\end{figure}

The free energy per baryon and the corresponding pressure at $T=50\ \mathrm{MeV}$ are shown in Fig.~\ref{fig2}. The triangle marks the minimum free energy per baryon, and the open circles correspond to the zero pressure. We notice that the pressure is zero when the density is at the lowest free energy per baryon, which is a necessary condition for thermodynamic consistency and consistent with Eq.~(\ref{30}).

\section{Evolution of strange quark stars}\label{SQSProp}
To describe the matter inside strange quark stars, leptons must also be taken into account.
Due to weak reactions such as  $d$,$ s  \leftrightarrow  u + e + \overline{\nu}_{e} $, $s + u \leftrightarrow  u + d $ and $e \leftrightarrow \mu + \nu_e + \overline{\nu}_{\mu}$, the chemical potentials $ \mu_{i}$ ($i = u, d, s, e$) need to satisfy weak equilibrium conditions
\begin{equation}\label{31}
\mu_{u}^{*}+\mu_{e}-\mu_{\nu_{e}}=\mu_{d}^{*}=\mu_{s}^{*},
\end{equation}
and
\begin{equation}\label{32}
\mu_{e}=\mu_{\mu}+\mu_{\nu_{e}}-\mu_{\nu_{\mu}}.
\end{equation}

The charge neutrality condition is
\begin{equation}\label{33}
2n_{u}-n_{d}-n_{s}-3n_{e}-3n_{\mu}=0,
\end{equation}
where the particle number densities are $n_i=n_{i}^{+}-n_{i}^{-}$.

In addition, the quark number density satisfies the baryon number conservation condition
\begin{equation}\label{34}
n_{b}=\sum_{q=u,d,s}\frac{1}{3}\left(n_{q}^{+}-n_{q}^{-}\right).
\end{equation}

This paper studies the evolution of properties of strange quark stars, including neutrino-trapping, deleptonization, neutrino-transparent, and the formation of a cold strange quark star.

Neutrinos are trapped in the early stages of a strange quark star with an unshocked and low entropy $(S/n_{b}\sim 1)$ core, known as the neutrino-trapped phase, during which the lepton fraction remains constant. The lepton fraction during the neutrino-trapped phase is
\begin{equation}\label{35}
Y_{L,e}=\frac{n_e+n_{\nu_{e}}}{n_b}=\xi,
\end{equation}
\begin{equation}\label{36}
Y_{L,\mu}=\frac{n_{\mu}+n_{\nu_{\mu}}}{n_b}=0,
\end{equation}
where $\xi \simeq 0.4$ depends on the efficiency of electron capture reactions in the early neutrino-trapped phase, which is the initial time $t\sim0\ \mathrm{s}$ before the supernova explosion~\cite{Prakash1997_PR280-1}. $Y_{L,\mu}=0$ indicates that there are no muons in matter when neutrinos are captured~\cite{Malfatti2019_PRC100-015803}.
\begin{figure*}[htbp]
\centering
\includegraphics[width=15cm]{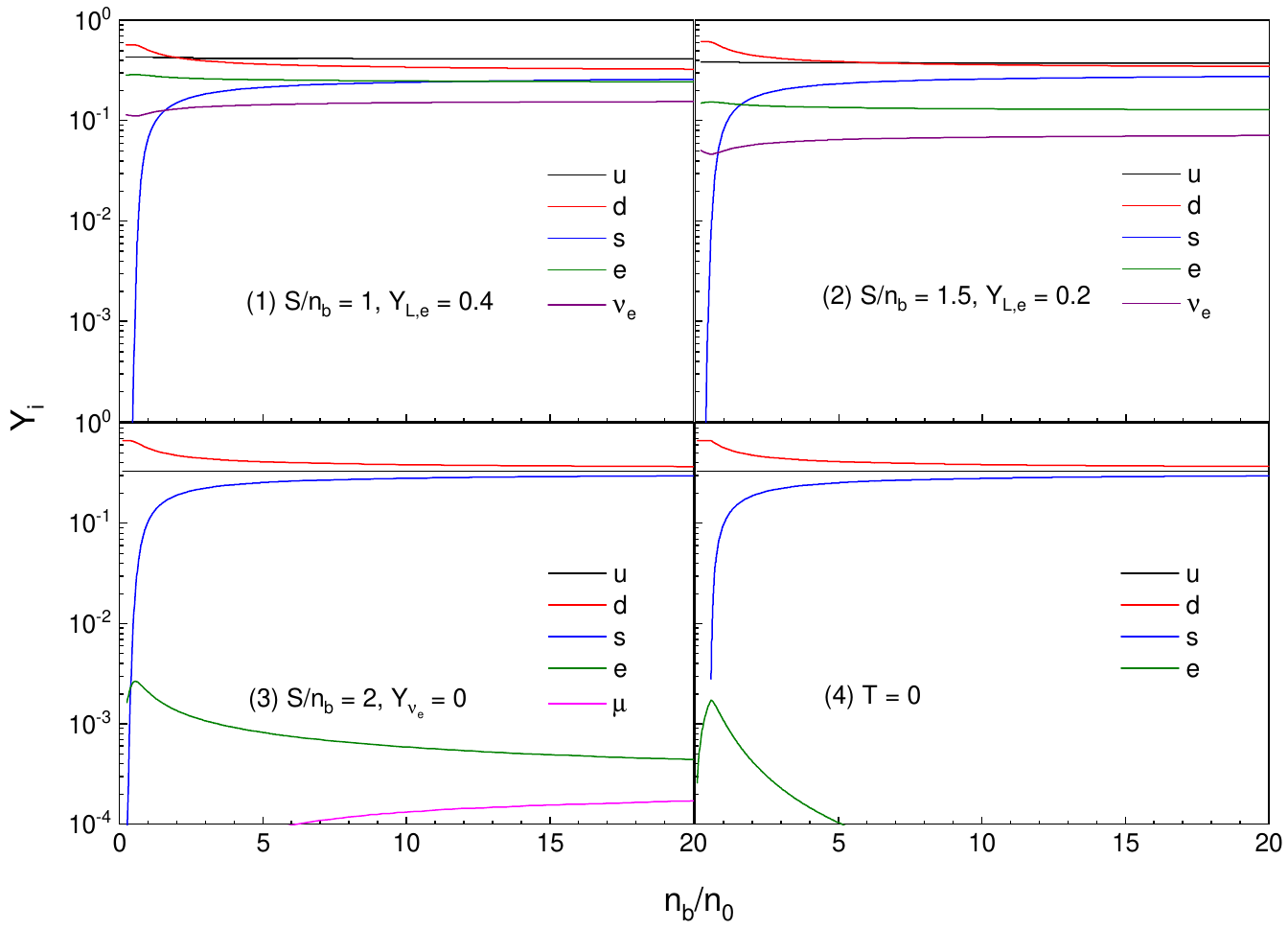}
\caption{The relative particle fraction of the matter inside of strange quark stars as functions of baryon density at various snapshots of its evolution. The upper and lower panels correspond to the neutrino-trapped and neutrino-transparent stages, respectively.}
\label{partifraction}
\end{figure*}

Then, approximately $0.5 - 1\ \mathrm{s}$ after the core bounce of a star, the explosion and shock wave of a stellar supernova lift off the stellar envelope. If a supernova explosion causes an extensive loss of neutrinos and deleptonization, leading to the extensive loss of lepton pressure, an increase in entropy per baryon, and mantle collapse, i.e., $S/n_{b}\sim 1.5$ and $\xi \simeq 0.2$, accretion becomes less important, and the star continues to evolve. If the intensity of the supernova explosion is not strong enough to cause the outer mantle of proto-strange quark stars to undergo deleptonization, the matter will gather again, leading to the formation of a black hole. This study considers the first scenario.

In the next $10 - 15\ \mathrm{s}$, the star undergoes neutrino diffusion, deleptonization, and core heating. The star begins to rapidly lose neutrinos, with a maximum entropy per baryon $S/n_{b}=2$ in the core and the neutrino fraction $Y_{\nu_e}$ reaching 0, and the equation of state will soften. If the gravitational pressure is large enough, a black hole could form. In this study, we assume that the strange quark star evolves to the final stage to form a cold-catalyzed strange quark star, and no black hole is formed during its evolution.

After deleptonization, the core of the hot strange quark star undergoes thermal diffusion and cooling. After about $50\ \mathrm{s}$, as the average energy of neutrinos decreases, the neutrinos in the star become essentially transparent, and the core reaches a cold-catalyzed configuration. Then the core continues to cool through thermal radiation, and finally, it shrinks to a cold-catalyzed strange quark star at $T=0 \ \mathrm{MeV}$.
\begin{figure}
\centering
\includegraphics[width=8cm]{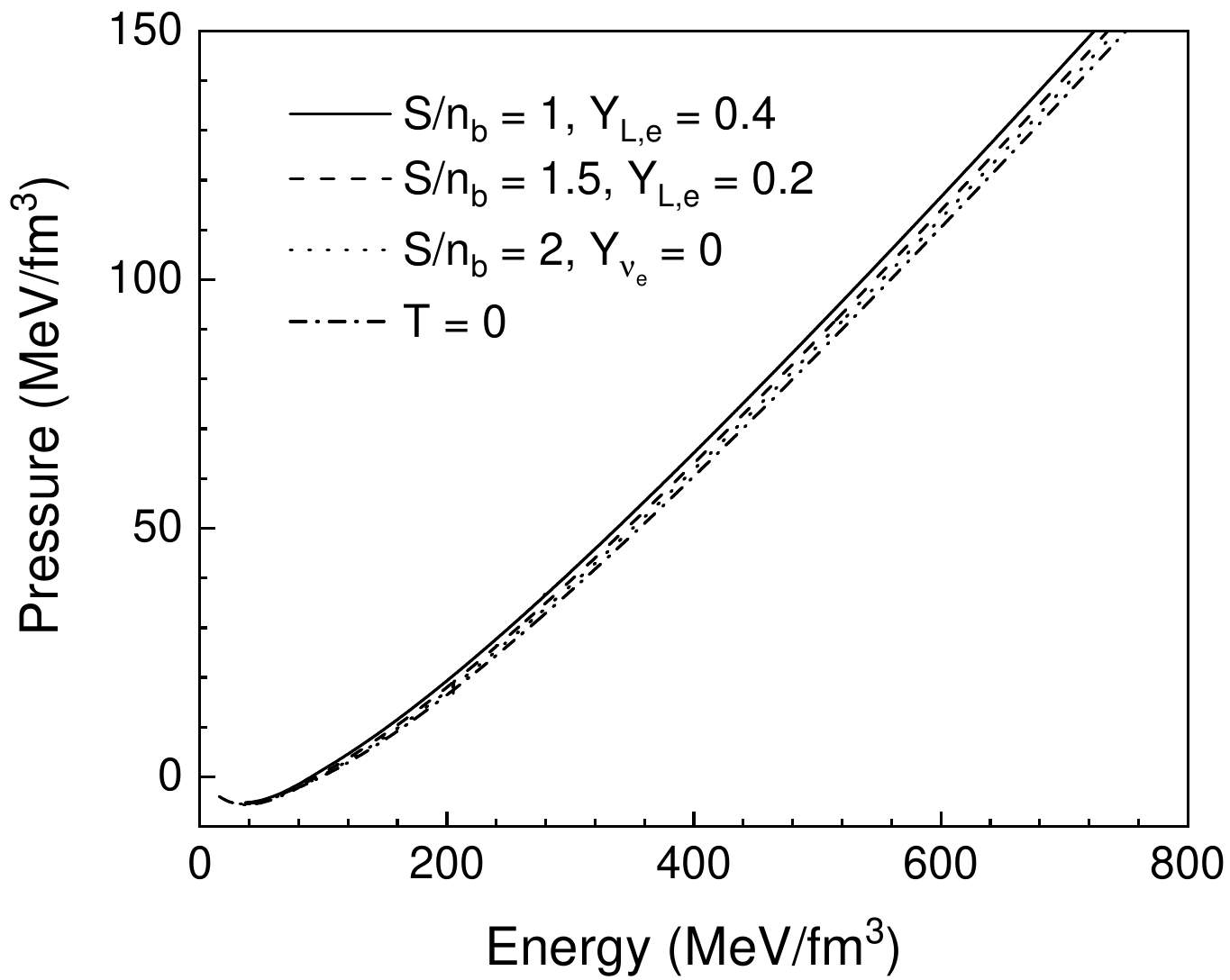}
\caption{The relationship between the pressure and energy density at different stages of evolution of SQS.}
\label{EOS}
\end{figure}

In Fig.~\ref{partifraction}, we show the relative fraction of particles of the matter inside strange quark stars as functions of baryon density at various snapshots along its evolution, from birth ($S/n_{b}=1, Y_{L,e}=0.4$) to cold-catalyzed strange quark stars at $T=0 \ \mathrm{MeV}$. The upper and lower panels correspond to the neutrino-trapped and neutrino-transparent stage, respectively. From the first stage to the second stage, we observe that higher neutrino concentrations increase the relative abundance of $u$ quarks while decreasing that of $d$ and $s$ quarks, leading to the emergence of $s$ quarks at higher $n_b$. In other words, the higher the neutrino concentrations, the lower abundances of $d$ and $s$ quarks, as well as the later appearance of $s$ quarks. This is consistent with some studies on the evolution of proto-neutron stars~\cite{Malfatti2019_PRC100-015803,Issifu2023_MNRAS522-3263}. For instance, research has shown that higher neutrino concentrations during the evolution of proto-neutron stars delay the emergence of particles with strangeness~\cite{Malfatti2019_PRC100-015803,Issifu2023_MNRAS522-3263}. Correspondingly, along the evolution lines of the star, the higher probabilities of $s$ quarks correlate with the softer EOS, meaning the lower maximum masses of the star. This will be validated in the research of mass-radius relation of strange quark stars. In addition, in the fourth stage, because the chemical potential is always smaller than the particle mass, the number density of $\mu$ is zero at $T=0 \ \mathrm{MeV}$.
\begin{figure}
\centering
\includegraphics[width=8cm]{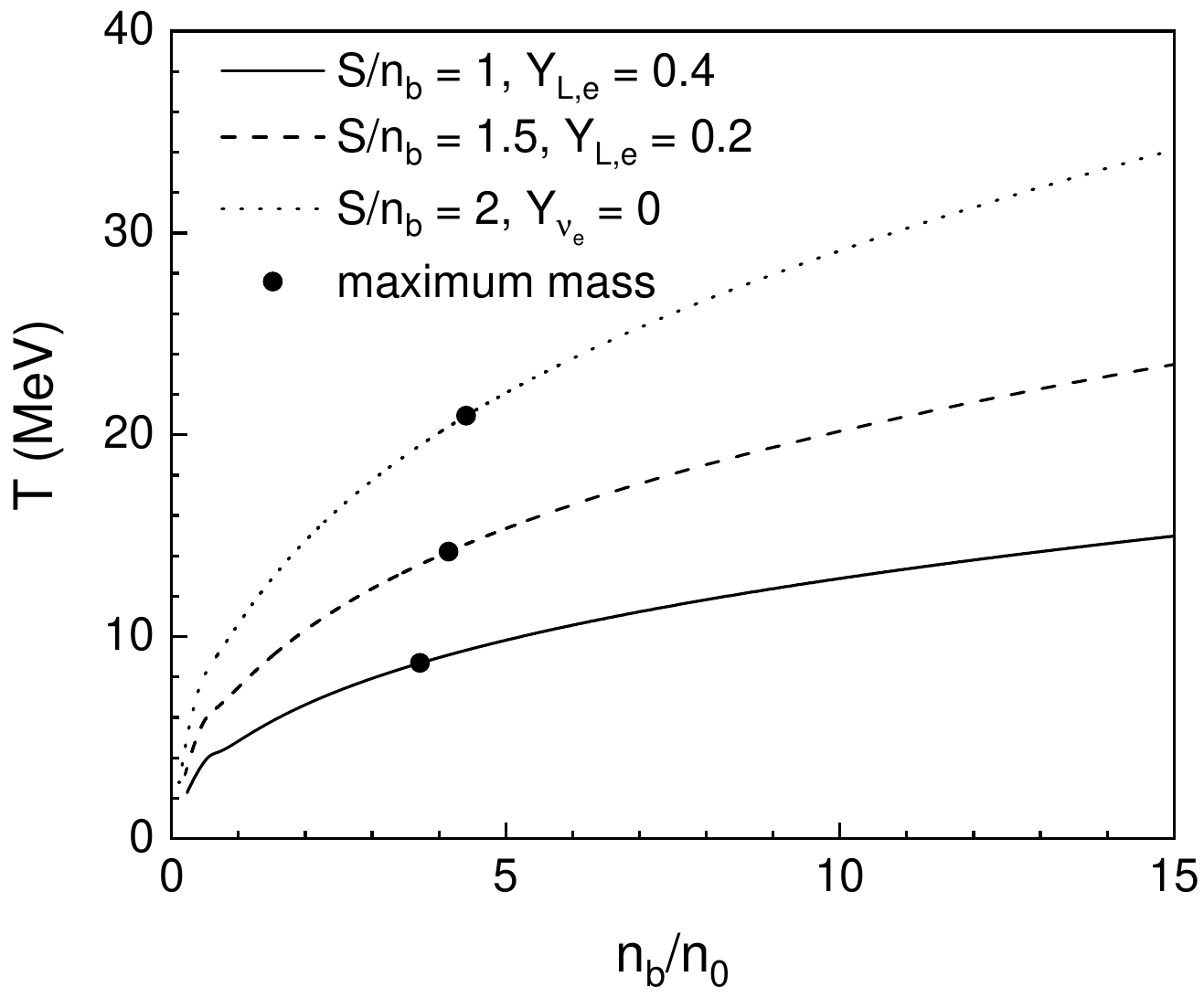}
\caption{The temperature as a function of baryon number density at different stages of evolution of SQS.}
\label{Temp}
\end{figure}

The equation of state of SQM at different stages of evolution of SQS is shown in Fig.~\ref{EOS}. It is clearly visible that the strange quark stars with lower neutrino concentrations exhibit a soft EoS, meaning lower pressure at the same energy density. This is because the probability of $s$ quarks appearing in the matter inside of strange quark stars increases with the deleptonization of evolution of star. At the same time, this also indicates that the maximum mass of a star will decrease when it evolves from the neutrino-trapped stage to the neutrino-transparent stage, which will be verified in the subsequent mass-radius relationship.

In Fig.~\ref{Temp}, we show the relationship between the temperature and baryon number density at different stages of evolution of SQS, where the black dots correspond to the core temperature of the matter inside strange quark stars. It can be seen that temperatures in the first phase are relatively low and below $15\ \mathrm{MeV}$ in most density ranges, which is the result of stellar collapse and core bounce. The baryon number density and temperature at the core of the most massive star are $3.72 n_{0}$ (the nuclear saturation density $n_0=0.152\ \mathrm{fm^{-3}}$) and $8.69\ \mathrm{MeV}$. In the second stage, the temperatures rises and remains below $23\ \mathrm{MeV}$ in most density ranges. This is due to neutrino emission and matter accretion, leading to deleptonization and an increase in the entropy density per baryon. The baryon number density and temperature at the core of the most massive star are $4.14 n_{0}$ and $14.21\ \mathrm{MeV}$. In the third stage, the temperature rises further and remains below $34\ \mathrm{MeV}$ in most density ranges due to the maximum heating of stellar matter by neutrino emission. The baryon number density and temperature at the core of the most massive star are $4.41 n_{0}$ and $20.94\ \mathrm{MeV}$. Subsequently, the core is continuously cooled by neutrino and thermal radiation and reaches $T=0\ \mathrm{MeV}$ after about 100 years. In addition, the temperature of a strange quark star is lower than that of its corresponding hadron star~\cite{Prakash1997_PR280-1,Issifu2023_MNRAS522-3263}, which allows us to determine the composition of compact objects by observing the evolution of surface temperature.
\begin{figure}
\centering
\includegraphics[width=8cm]{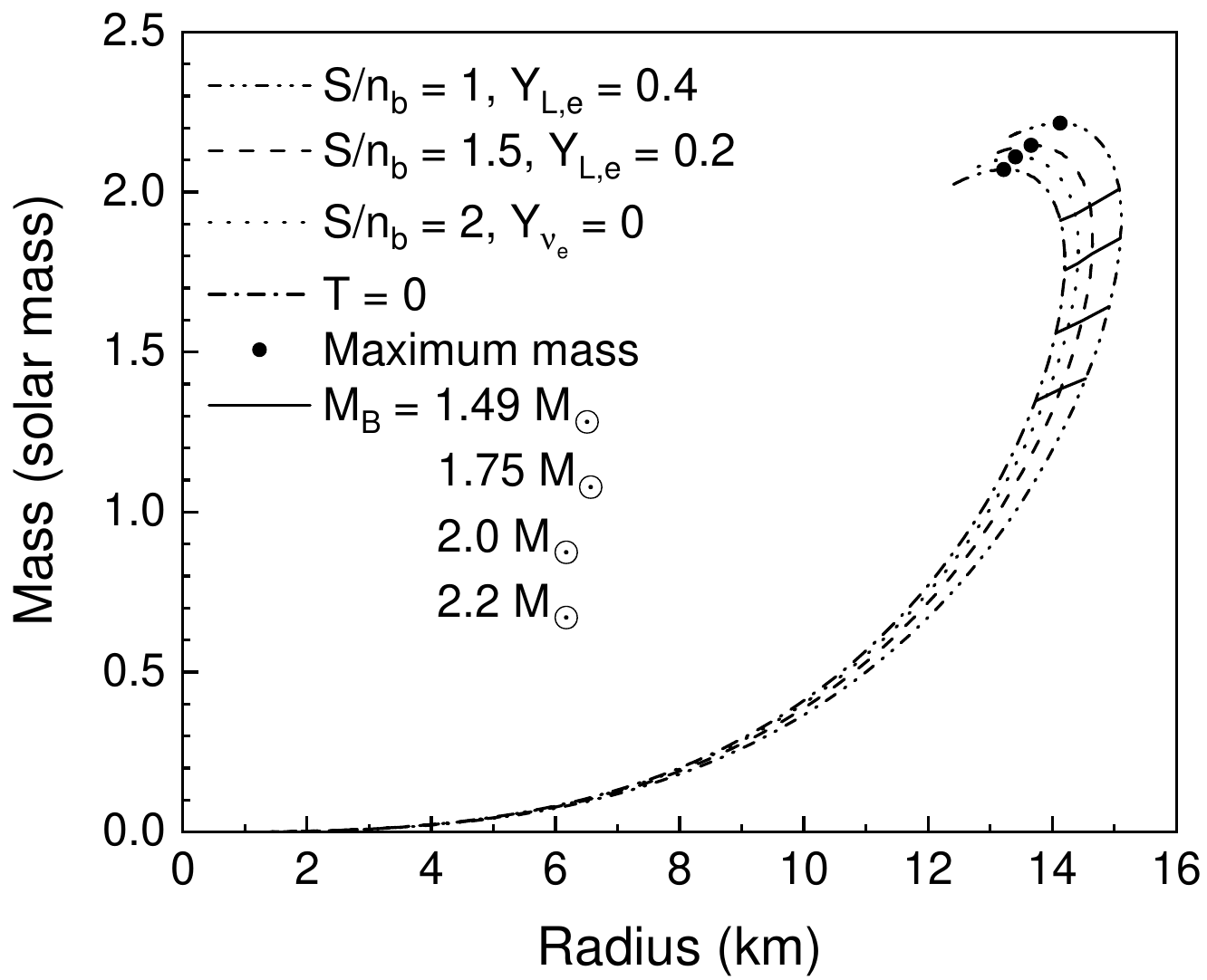}
\caption{The mass-radius relationship for strange quark stars at different stages of evolution of SQS.}
\label{MR}
\end{figure}
\begin{table}
\centering
\caption{
\label{tabLam}
The maximum mass and its corresponding radii, the central baryon density, and core temperature of stars for strange quark stars at different stages of evolution of SQS. }
\begin{tabular}{ccccccccccc}\hline
$S/n_{B}$ && $Y_{L,e}$ && $R$\mbox{(km)} && $M_{max}$\mbox{($M_{\odot}$)} && $n_{c}$\mbox{($n_{0}$)} && $T_{c}$\mbox{(MeV)} \\
\hline
1.0     &&0.4            &&14.13 &&2.21 &&3.72 &&8.69 \\
1.5     &&0.2            &&13.66 &&2.15 &&4.14 &&14.21 \\
2.0     &&$Y_{\nu,,e}=0$ &&13.41 &&2.11 &&4.41 &&20.94 \\
$T=0$   &&$Y_{\nu,,e}=0$ &&13.22 &&2.07 &&4.54 &&- \\
  \hline
\end{tabular}
\end{table}
\begin{figure}
\centering
\includegraphics[width=8cm]{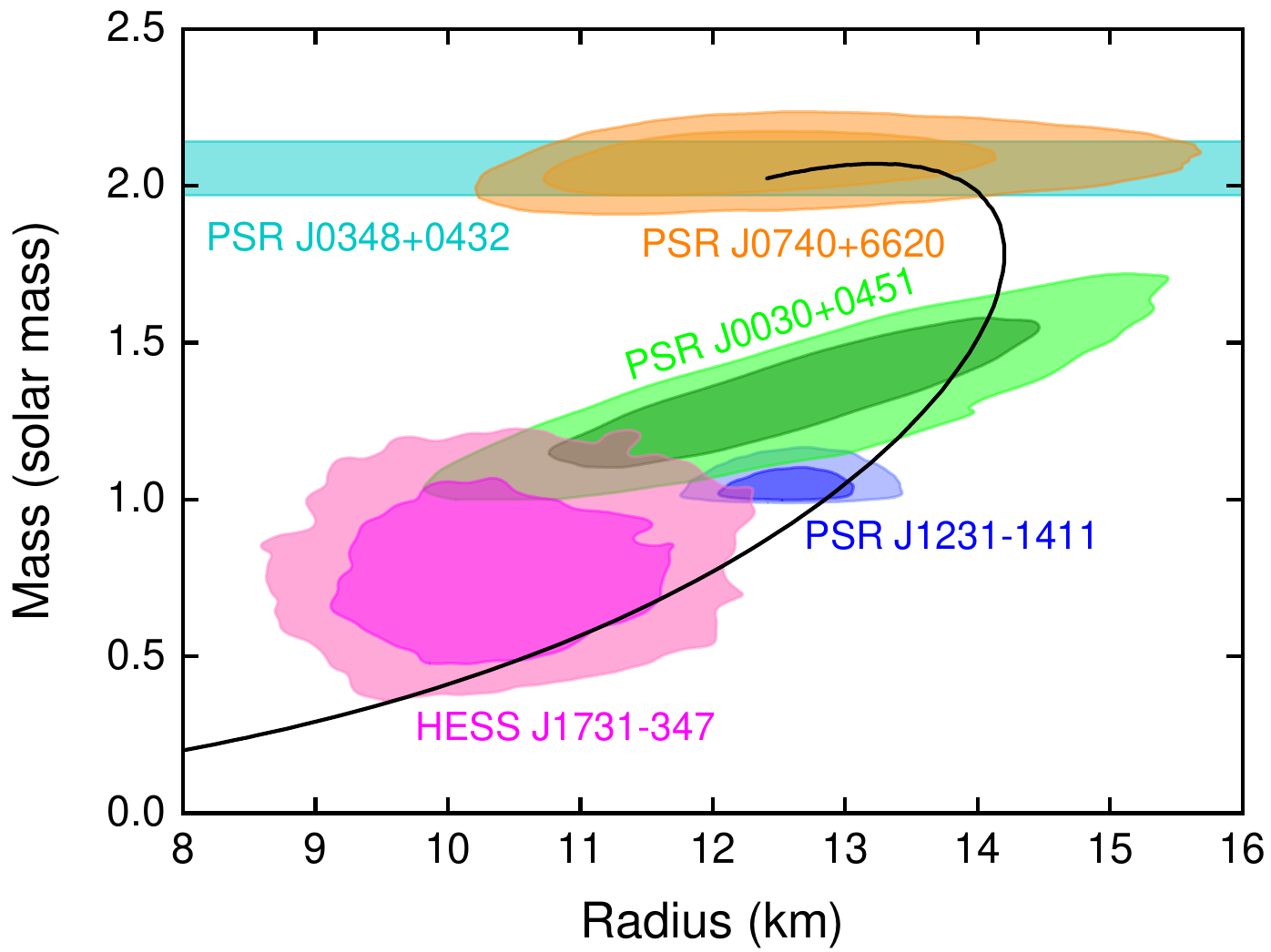}
\caption{The mass-radius relationship for strange quark stars at $T=0$, where $D^{1/ 2}=130.9\ \mathrm{MeV}$ and $C=0.69$.}
\label{MR1}
\end{figure}

Based on the EOSs, we solve the Tolman-Oppenheimer-Volkov equation
\begin{equation}
\frac{\mathrm{d}P}{\mathrm{d}r}=-\frac{GmE}{r^{2}}\frac{(1+P/E)(1+4\pi r^{3}P/m)}{1-2Gm/r},
\end{equation}
and
\begin{equation}
\mathrm{d}m=4\pi Er^{2}\mathrm{d}r.
\end{equation}
\begin{figure*}[htbp]
\centering
\includegraphics[width=15cm]{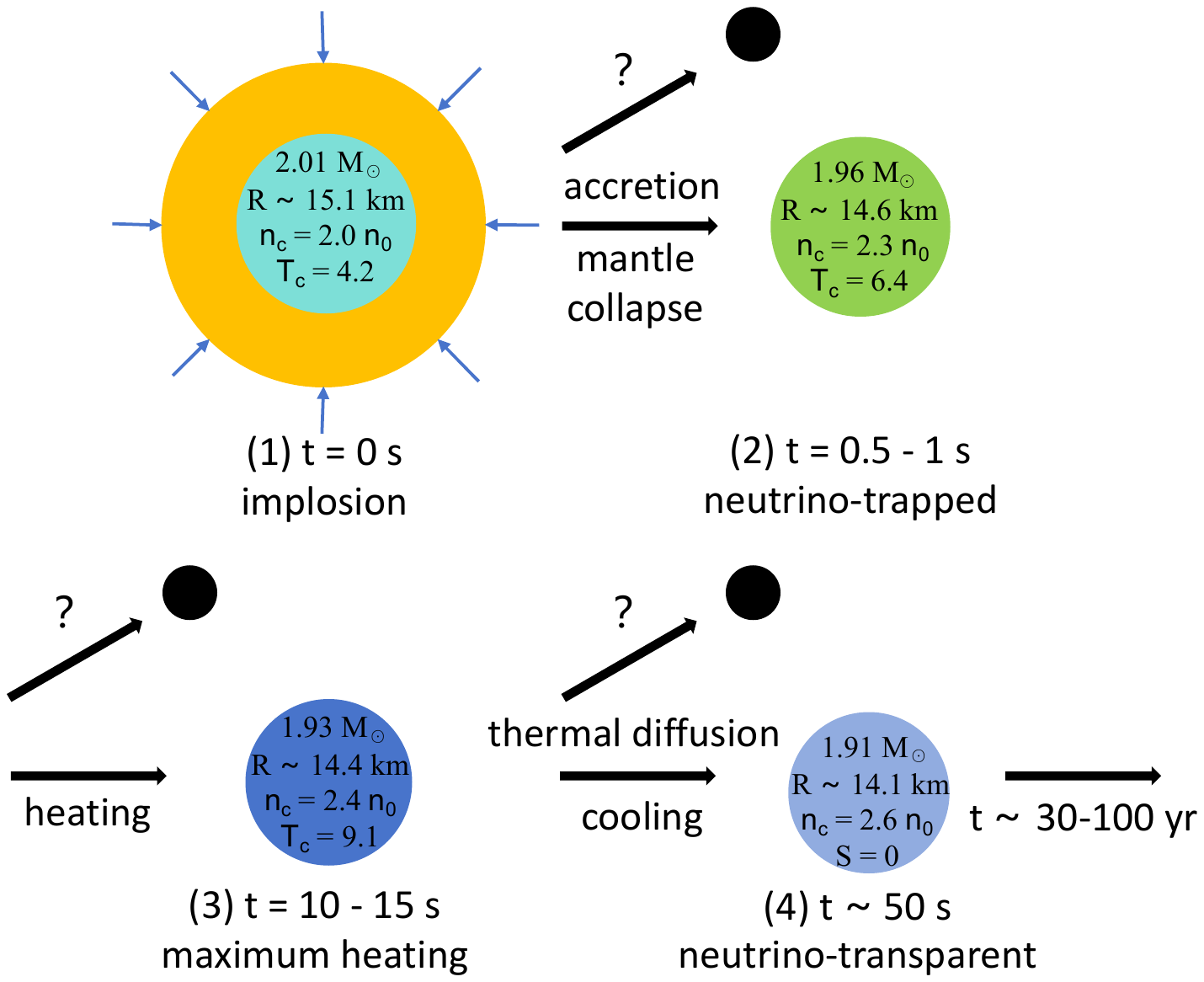}
\caption{The schematic diagram of the state of different temporal stages in the evolution from proto-strange quark star to a cold strange quark stars at fixed baryon-mass $M_{B}=2.2\ M_{\odot}$.}
\label{diagram}
\end{figure*}

The mass-radius relation for strange quark stars at different stages of evolution of SQS is shown in Fig.~\ref{MR}, where the most massive stars are represented by black dots.
We can see that the maximum mass and corresponding radius of strange quark stars gradually decrease when they evolve from the neutrino-trapped stage to the neutrino-transparent stage. To clearly illustrate the states during the evolution process, the maximum mass and its corresponding radii, the central baryon density, and core temperature of stars for strange quark stars at different stages of evolution of SQS, are listed in Tab.~\ref{tabLam}. We can observe that the maximum mass and its corresponding radii continuously decreases, and the central baryon density and core temperature continuously rises during the evolution process. The maximum mass of strange quark stars in each stage exceeds $2\ M_{\odot}$. The decrease in the maximum mass of strange quark star is consistent with the softening of the equation of state in Fig.~\ref{EOS} as it evolves. In addition, we also presented strange quark stars with baryon masses $M_B=2.2,2.0, 1.75, 1.49 \ M_{\odot}$ in each stage, and found that the evolution of strange quark star with constant baryon masses is approximately shown as a straight line in the figure.

The mass-radius relationship for strange quark stars at $T=0$ is shown in Fig.~\ref{MR1}. We have also included 2-D marginalized posteriors of masses and radii of HESS J1731-347~\cite{Doroshenko2022NA6}, PSR J0030+0451~\cite{TER2019AJ887}, PSR J1231-1411~\cite{Salmi2024AJ976,Qi2025AJ981}, and PSR J0740+6620~\cite{Riley2021AJL918}. The contours in the 2-D marginalized posterior denote the $68\%$ and $95\%$ credible intervals. In addition, the lower and upper limits of the mass and radius of PSR J0348+0432~\cite{JA2013Sci340} were also included. It can be seen that the results of the strange quark stars with $C=0.69$ and $D^{1/2}=130.9\ \mathrm{MeV}$ are consistent with the observational masses and radii of HESS J1731-347, PSR J1231-1411, PSR J0030+0451, PSR J0348+0432, and PSR J0740+6620. The most interesting aspect of the SQS model is that it describes compact stars with masses up to approximately $1.5\ M_{\odot}$ and radii between 8 and $12\ \mathrm{km}$, such as PSR J0030+0451 and HESS J1731-347, which can explain the observation results of some cold, dense, and small compact stars that do not conform to the standard NS model. A recent study also revealed that quark-star models naturally accommodate the ultra-low-mass object HESS J1731-347, a configuration that is difficult to reproduce in standard gravity-bound neutron-star models~\cite{Xie20266_PRD114-023016}.

Based on the above analysis, we provide a schematic diagram in Fig.~\ref{diagram}, illustrating the state of different temporal stages in the evolution of proto-strange quark stars to strange quark stars at fixed baryon mass. The baryon mass is given by~\cite{Malfatti2019_PRC100-015803}
\begin{equation}
M_{B}=m_{n}\int_{0}^{R}\frac{4\pi r^{2}n(r)}{[1-2Gm(r)/r]^{1/2}}\mathrm{d}r.
\end{equation}
At fixed $M_{B}=2.2\ M_{\odot}$, the gravitational mass and radius decrease, while the baryon density and temperature at the core increase as the star evolves, ultimately becoming a stable cold strange quark star.

\section{Summary}\label{Summ}
We studied the properties of strange quark matter and revealed the evolution process of strange quark stars by using a self-consistent thermodynamic treatment method.

It is found that higher neutrino concentrations in the early stage increase the relative abundance of $u$ quarks while decreasing that of $d$ and $s$ quarks, leading to the emergence of $s$ quarks at higher $n_b$. This is consistent with some studies on the evolution of proto-neutron stars. Based on the equation of state of the matter inside strange quark stars, it can be seen that the strange quark stars with lower neutrino concentrations correspond to a soft EoS because the probability of $s$ quarks increases, meaning the maximum mass of a star will decrease when it evolves from the neutrino-trapped stage to the neutrino-transparent stage.
The thermal profile of the baryon number density is also examined through three distinct evolutionary stages. During the initial stage, temperatures are constrained below 15 MeV over most of the density range, attributable to the dynamics of stellar collapse and core bounce. The second stage is characterized by a moderate increase to below 23 MeV, primarily driven by neutrino emission and ongoing mass accretion, which promote deleptonization and a rise in entropy per baryon. The third stage exhibits further warming to below 34 MeV, resulting from maximum heating of stellar matter by neutrino emission. Thereafter, the core undergoes progressive cooling over roughly a century, driven by neutrino and thermal radiation, ultimately relaxing to $T=0\ \mathrm{MeV}$. Furthermore, the systematically lower temperatures predicted for strange quark stars compared to hadronic stars, which allows us to determine the composition of compact objects by observing the temperature evolution.

Based on the EOSs, we obtained the mass-radius relationships of strange quark stars at different stages of evolution of SQS. It can be seen that the maximum mass and corresponding radius of strange quark stars gradually decrease along the evolution of stars. The maximum mass and corresponding radius of stars are $2.21\ M_{\odot}$ and $14.13\ \mathrm{km}$ at the first stage, the maximum mass and corresponding radius of strange quark stars are $2.15\ M_{\odot}$ and $13.66\ \mathrm{km}$ at the second stage, the maximum mass and corresponding radius of strange quark stars are $2.11\ M_{\odot}$ and $13.41\ \mathrm{km}$ at the third stage, and the maximum mass and corresponding radius of stable strange quark stars are $2.07\ M_{\odot}$ and $13.22\ \mathrm{km}$ at zero temperature. The maximum mass of strange quark stars in each stage exceeds $2\ M_{\odot}$. The decrease in the maximum mass of strange quark star is consistent with the softening of the equation of state as it evolves.
In addition, we found that the mass-radius relationship of the cold strange quark star with $C=0.69$ and $D^{1/2}=130.9\ \mathrm{MeV}$ is consistent with the observational masses and radii of HESS J1731-347, PSR J1231-1411, PSR J0030+0451, PSR J0348+0432, and PSR J0740+6620, which can explain the observation results of some cold, dense, and small compact stars that do not conform to the standard NS model.
Finally, we provide a schematic diagram illustrating the state of different stages along the evolution of stars at fixed baryon mass. It is found that the gravitational mass and radius decrease, while the baryon density and temperature at the core increase as the star evolves, ultimately becoming a stable cold strange quark star.\par

\section*{ACKNOWLEDGMENTS}

The authors would like to thank support from NSFC (Nos. 11135011, 12275234, 12375127), the Natural Science Foundation of Fujian Province, China (No. 2024J01319), the Young and Middle-aged Teachers' Educational Research Project of Fujian Province, China (No. JAT231123), and the Scientific Research Project of Wuyi University, China (No. YJ202212).

\end{document}